\documentclass[12pt]{article}

\usepackage{amscd}

\usepackage{verbatim}

\usepackage{amssymb}

\usepackage{amsmath}

\newcommand{\be}{\begin{equation}}
\newcommand{\ee}{\end{equation}}
\newcommand{\ben}{\begin{eqnarray}}
\newcommand{\een}{\end{eqnarray}}
\newcommand{\ba}{\begin{eqnarray}}
\newcommand{\ea}{\end{eqnarray}}

\newcommand{\bi}{\begin{itemize}}
\newcommand{\ei}{\end{itemize}}

\begin{document}

\begin{center}

\vspace{24pt} { \large \bf Necessary conditions for an AdS-type instability} \\

\vspace{30pt}

\vspace{30pt}

\vspace{30pt}

{\bf Dhanya S. Menon\footnote{dhanya.menon@students.iiserpune.ac.in}}, {\bf Vardarajan
Suneeta\footnote{suneeta@iiserpune.ac.in}}

\vspace{24pt} 
{\em  The Indian Institute of Science Education and Research (IISER),\\
Pune, India - 411008.}

\end{center}
\date{\today}
\bigskip

\begin{center}
{\bf Abstract}
\end{center}
In this work, we analyze the necessary conditions for a nonlinear AdS-type instability in the gravity-scalar field system. In particular, we discuss the necessary conditions for a cascade of energy to higher modes by applying results in KAM theory. Our analytical framework explains numerical observations of instability of spacetimes even when the spectrum is only asymptotically resonant, and the fact that a minimum field amplitude is needed to trigger it. Our framework can be applied for similar stability analyses of fields in (locally) asymptotically AdS spacetimes. We illustrate this with the example of the AdS soliton. Further, we conjecture on the possible reasons for quasi-periodic behaviour observed in perturbations of AdS spacetime.  Certain properties of the eigenfunctions of the linear system dictate whether there will be localization in space leading to black hole formation. These properties are examined using the asymptotics of Jacobi polynomials. Finally we discuss recent results on the AdS instability in Einstein-Gauss-Bonnet gravity in light of our work.

\newpage
\section{Introduction} Anti-de Sitter ($AdS$) spacetime is the arena for the $AdS/CFT$ conjecture that provides a dual description of gravity in (an aymptotically) $AdS$ spacetime (with other fields) in terms of a non-gravitational field theory on its boundary. In this context, there is a huge body of work on field propagation in $AdS$ spacetime treated in a linearized approximation. The importance of nonlinearities in $AdS$ has since been realized due to recent work by Bizon and Rostworowski \cite{bizon} that demonstrates that $AdS$ spacetime is nonlinearly unstable to massless scalar field perturbations, however small. Their result is primarily numerical, and the end-point of the instability is the formation of a black hole in four dimensions and higher (see also \cite{lecture}).\footnote{ In three dimensions, a black hole does not form, but nevertheless, initially smooth perturbations lose smoothness, and their higher Sobolev norms grow with time \cite{jalmuzna1}, \cite{jalmuzna2}.} They have demonstrated the nature of the instability by considering weakly nonlinear perturbations. The nonlinearity triggers a resonant instability wherein energy is transferred to higher frequency modes (of the linearized system). A similar result holds for complex scalar fields \cite{buchel}.

In this paper, we investigate the necessary conditions for (i) an $AdS$-like (resonant) instability to occur in field propagation in spacetimes, and (ii) given such an instability, the necessary condition for it to result in black hole formation.
An aim is to pin down what features of field propagation in $AdS$ such as a perfectly resonant spectrum of linear modes, or reflecting boundary conditions are crucial for the resonant instability. A more important reason is to investigate if asymptotically $AdS$ spacetimes suffer from nonlinear instabilities as $AdS$ does. An obvious application of this is to the $AdS/CFT$ correspondence.

In section II, we show that a paradigm shift in viewing the $AdS$ instability is the key to understanding when resonant instability can occur in the gravity-field system. As we discuss in II.2, perturbation theory in $AdS$ can be studied as perturbations of an appropriate integrable Hamiltonian --- integrability implies dynamics is restricted to be on tori in phase space, and instability due to the perturbation leads to destruction of these tori. The $AdS$ instability is thus an instability in phase space. In II.3, we apply a theorem in nonlinear dynamics due to Benettin and Gallavotti \cite{bg} which is specific to the integrable Hamiltonian that emerges in the gravity-scalar field system. Using this, we derive the necessary conditions for these tori to be destroyed, and these in turn lead to conditions under which a instability in phase space can happen. If the operator governing linear perturbations is self-adjoint with the chosen boundary conditions, then the necessary condition for instability is a linear spectrum that is discrete and either resonant or approximately resonant (in a precise sense). In the approximately resonant case, a minimum nonzero field amplitude (that depends on the deviation of the spectrum from perfect resonance) is needed to trigger the instability. An example is an asymptotically resonant spectrum (approaching resonant for high mode number) which arises in fields in a cavity in Minkowski spacetime with Neumann boundary conditions. Our analysis explains for the first time, the puzzling numerical studies which show in this case that the turbulent energy cascade can occur \cite{maliborski2}, but only with a minimum nonzero scalar field amplitude \cite{MR} (see also \cite{pani}). In II.4 (and Appendix A), we demonstrate the application of our analysis to (locally) asymptotically $AdS$ spacetimes by considering scalar field perturbations of the $AdS$ soliton in weakly nonlinear perturbation theory. We identify a class of metric perturbations of the $AdS$ soliton caused by the back-reaction of the field, for which the equations are very similar to the $AdS$ scalar field system. Using our framework, we derive the necessary conditions for instabilities in this system. Further nonlinear aspects are discussed in section II.5.

For the end-point of the instability to be a black hole, we also need a sufficient localization of energy in space. Given a resonant instability that transfers energy to higher modes, in section III, we argue that a necessary condition for localization of energy is localization of the high mode linear eigenfunctions, at least for small field amplitudes. This is demonstrated explicitly for $AdS$, where the asymptotics (high mode behaviour) of the mode functions (Jacobi polynomials) is studied using the Mehler-Heine formula for various dimensions. Localization of higher mode functions is least for $AdS_3$ consistent with the observation in \cite{jalmuzna1} that black hole formation does not occur in $AdS_3$ despite a very fast transfer of energy to higher modes (in comparison with higher dimensions). This leads to an understanding of when an energy cascade can lead to black hole formation in an asymptotically $AdS$ spacetime. Section II and III together derive necessary conditions for an $AdS$-like  nonlinear (in)stability in asymptotically $AdS$ spacetimes --- this can now be checked by a simple investigation of the spectrum and eigenfunctions of the linear problem.

Bizon and Rostworwski's work has led to several efforts to obtain a better analytical understanding of the instability. Among these has been the observation \cite{balasubramanian} that the $AdS$ instability is similar to the evolution of a system of coupled nonlinear oscillators first studied by Fermi, Pasta, Ulam and Tsingou (FPUT) \cite{FPU}. Similar to the FPUT problem, the authors of \cite{balasubramanian} separate the fast and slow time scales in the system. A careful analysis of this separation using rigorous renormalization group methods was done in \cite{craps1}. The two-time framework has been used since in \cite{craps}, \cite{bizon2}. This has uncovered quasi-periodic and recurrent behaviour in the gravity-scalar field system for times $O(1/\epsilon^2)$, $\epsilon$ being the field amplitude at initial time. Our work, on the other hand, gives predictions about the behaviour of the system for exponentially long times in $1/\epsilon^2 $. We also conjecture on reasons for this quasi-periodic behaviour in section II.5. Another area of investigation in this topic is whether special initial perturbations are stable. In particular, fine-tuned periodic perturbations which were constructed in \cite{maliborski} and boson stars seem to be stable \cite{buchel, buchel1} (see also \cite{maliborski1}). These are the nonlinear versions of single modes of the linearized analysis (`oscillons'). For evolution of initial perturbations comprising of two modes of the linearized analysis, numerical data is inconclusive, but this has led to lively debate \cite{bizon1}, \cite{buchel2}, and a more recent numerical analysis is in \cite{frey2}. It has been surmised that there may be quasi-periodic solutions consisting of many modes for special initial conditions \cite{frey2}. The issue of whether gravitational interaction localizes energy in position space has been investigated in \cite{freivogel} by using shells of fields. Our work adopts a different approach --- of studying asymptotics of eigenfunctions which explains the dimension dependence of localization, such as no horizon formation in $AdS_3 $.

The results of our paper are summarized in section IV. In section V, we have an extended discussion of open problems in this area. In particular, we discuss a recent result in Einstein-Gauss-Bonnet gravity coupled to a scalar field of amplitude $\epsilon$ \cite{kunstatter} which shows complex behaviour in comparison to the pure gravity case. We argue that under a field rescaling these equations are identical to the pure gravity case to $O(\epsilon^3)$, but not beyond. The details of this computation are given in Appendix B.

\section{Necessary conditions for $AdS$-type resonant instability in gravity-scalar field system}
In this section, we obtain a deeper understanding of the $AdS$ instability using nonlinear dynamics. This paradigm shift enables us to understand precisely when field perturbations in a spacetime can lead to a resonant instability with a cascade of energy to higher frequencies as observed in $AdS$. We obtain necessary conditions for such a resonant instability in the gravity-scalar field system, and this analysis can be directly applied to scalar (and other) fields in asymptotically $AdS$ spacetimes.

\subsection{Summary of gravity-scalar field set-up in $AdS$}
We first summarize the Einstein gravity-scalar field set-up used by Bizon and Rostworowski to uncover the instability of $AdS$ spacetime \cite{bizon}. Their first paper considered $(3+1)$ dimensions, but for future use, we write the equations in $(d+1)$-dimensions. Let $\Lambda = \frac{-d(d-1)}{2 l^2 }.$
The coupled field equations are:
\begin{equation}
G_{ab}+\Lambda g_{ab}=8\pi G \left( \partial_a \phi \partial_b\phi-\frac{1}{2}g_{ab}(\partial\phi)^2 \right ) .
\label{dp1}
\end{equation}

\begin{equation}
g^{ab}\nabla_a \nabla_b\phi=0.
\label{dp2}
\end{equation}
The scalar field and metric are assumed to have spherical symmetry with the metric ansatz being
\begin{equation}
ds^2=\frac{l^2}{\cos^2x}(-Ae^{-2\delta}dt^2+A^{-1}dx^2+\sin^2x d\Omega^{2}_{d-1}) .
\label{dp3}
\end{equation}
where $d\Omega^{2}_{d-1}$ is the standard metric on the $(d-1)$-sphere. The range of coordinates are $-\infty<t<\infty$ and $0\leq x <\pi/2$. $A$, $\delta$ and $\phi$ are functions of $t$ and $x$ only. The pure $AdS$ solution is given by $A=1$, $\delta=0$ and $\phi=0$. In this system, reflecting boundary conditions are imposed on the scalar field.\footnote{For technical subtleties associated to the equations of the gravity-scalar field system with this boundary condition, see \cite{friedrich}.}

A solution to the Einstein-scalar field system in weakly nonlinear perturbation theory is obtained by the expansion in a small parameter $\epsilon >0$ as follows (to make contact with previous literature, we use units: $8\pi G = d-1;~ c=1$):
\begin{eqnarray}\phi=\sum\limits_{j=0}^\infty\phi_{2j+1}\epsilon^{2j+1};~~A=1-\sum\limits_{j=1}^\infty A_{2j}\epsilon^{2j} ; ~~\delta=\sum\limits_{j=1}^\infty\delta_{2j}\epsilon^{2j} .\label{dp10}\end{eqnarray}
The zeroth order (in $\epsilon$) background metric is $AdS$ with $\phi=0$.
The first-order (in $\epsilon$) equation is $\ddot{\phi_1}+ L\phi_1=0 $ where
\begin{equation} L = - \frac{1}{\tan^{d-1} x}\partial_x \left[(\tan^{d-1} x ) \partial_x \right]. \label{dp11} \end{equation}
The solution obeying reflecting boundary conditions at the $AdS$ boundary $x = \pi/2$ is of the form $\sum\limits_{j} \phi_{1j} = \sum\limits_{j} a_j \cos(\omega_j t+\beta_j) e_j(x)$.
$e_j(x)$ are the eigenfunctions of the linear operator $L$ with eigenvalues $\omega_{j}^{2}= (2j+d)^{2}$. $L$ is self-adjoint on the space of square integrable functions with inner product $$<f,g> = \int_{0}^{\pi/2} f(x) g(x) \tan^{d-1} x~dx .$$ The (orthogonal) eigenfunctions are
\begin{equation}
e_j(x)=\frac{2\sqrt{j!(j+d-1)!}}{\Gamma(j+d/2)}\cos^dx P_j^{\frac{d}{2}-1,\frac{d}{2}}(\cos2x);
\label{dp12}
\end{equation}
where $P_j^{\alpha,\beta}$ are the Jacobi polynomials and $j$ are non-negative integers.

At order $\epsilon^2$, as a result of the back-reaction of the scalar field, we have:
\begin{equation}\delta_2=-\int_0^x\cos(y)\sin(y)[(\phi_1')^2+(\dot{\phi_1})^2 ]dy .\label{dp15}\end{equation}
\begin{equation}A_2=\frac{\sin x \cos x}{\tan^{d-1} x}\int_0^x [ (\phi_1')^2+(\dot{\phi_1})^2 ] \tan^{d-1} y ~dy .\label{dp16}\end{equation}
At the third order an inhomogeneous equation is obtained
\begin{eqnarray}\ddot{\phi_3}+ L\phi_3=S(A_2,\delta_2,\phi_1)
=- 2\ddot{\phi_1}(\delta_2+A_2) - \dot{\phi_1}(\dot{\delta_2}+\dot{A_2}) - \phi_1'(A_2'+\delta_2').~~~
\label{dp16a}\end{eqnarray}
We can expand $\phi_3 = \sum\limits_{j} c_j (t)e_j (x)$ as $\{e_j (x) \}$ form a basis on the space of functions on which $L$ is self-adjoint (assuming $\phi_3$ lies in this function space).
Projecting (\ref{dp16a}) onto the basis $\{ e_j (x) \}$ results in
\begin{equation}
\ddot{c_j}+\omega_j^2c_j=(S,e_j(x))
\label{dp17}
\end{equation}
which is an infinite set of forced harmonic oscillator equations for the coefficients $c_j(t)=(\phi_3,e_j(x))$.
Triads of frequencies in $(S,e_j(x))$  of the form $\{\omega_{j_1}, \omega_{j_2}, \omega_{j_2} \}$ can lead to resonances when $\pm \omega_j=\omega_{j_{1}}\pm \omega_{j_{2}}\pm \omega_{j_3}$. This is due to the perfectly resonant linear spectrum of the scalar field in $AdS$. Not all secular terms can be removed, leading to a resonant instability and transfer of energy to high frequency modes over time.

\subsection{Instability of the scalar field-gravity system from the nonlinear dynamics point of view}
We now characterize the resonant instability in a different way, which yields many benefits.
Consider the equation for the scalar field (\ref{dp2}), with the metric (\ref{dp3}) solving the Einstein equation (\ref{dp1}). The computations in the previous section show that we can expand the scalar field as $\sum\limits_{j} \tilde a_{j}(t) e_{j}(x)$ in terms of the basis $\{ e_{j}(x) \}$ of eigenfunctions of a self-adjoint operator.\footnote{ Existence results for the spherically symmetric Einstein-scalar field system \cite{holzegel}, \cite{holzegel1} suggest that in the case of  \emph{even} spacetime dimensions, the basis for the scalar field of the full nonlinear system may be a corrected version of the linear eigenfunctions with corrections being higher order in $\epsilon$. This will not change the discussion of this subsection and only add one more term to the source in (\ref{dp17}). } Then, in weakly nonlinear perturbation theory (\ref{dp10}), $\tilde a_{j}(t)$ can be expanded as a series of odd powers of $\epsilon$. At order $\epsilon$, the $\tilde a_{j}(t)$ are just harmonic, with frequencies $\omega_j $. The next (third) order correction is given by solutions of a harmonic oscillator equation with forcing terms from the nonlinearities, as in (\ref{dp17}). The equations governing $\tilde a_j (t)$ thus arise from a Hamiltonian that can be viewed as a perturbation of linear harmonic oscillators, where the perturbation couples modes, leading to forcing terms at higher orders. In principle, the gravity-scalar field system has an infinite number of such coupled nonlinear oscillators. But at any finite time (however large), we may expect significant energy transfer to only a finite number of modes. In studies of the $AdS$ instability, most numerical simulations show a decaying power law behaviour of energy per mode with mode frequency, close to black hole formation times \cite{lecture}. In a two-time formalism, conservation laws have been found \cite{craps} (which may hold at times of the order $\frac{1}{\epsilon^2}$) that lead to inverse energy cascades \cite{lehner}. Thus we may view the Hamiltonian for $\tilde a_j (t)$ as a perturbed system of $n$ linear (undamped) harmonic oscillators, where $n$ can be large but is finite. As we shall see, the analysis that follows from this explains puzzling numerical results in the literature.

The Hamiltonian for $\tilde a_j (t)$ can be written as an \emph{integrable} part ($n$ linear harmonic oscillators) plus a perturbation of order $\epsilon^2$. By the Liouville-Arnold theorem, the integrable system of $n$ linear oscillators can be described in terms of action-angle variables $\{(I_{i}, \theta_{i})\} \in R^{n} \times T^{n} $, $i = 1,...,n$ and dynamics takes place on $n$-tori, which are level surfaces of the action variables. We use the notation $(I, \theta)$ to collectively describe the $2n$ variables, dropping subscripts. Denoting the Hamiltonian of the integrable $n$ linear oscillators by $ h_{0}(I)$, the Hamiltonian of the perturbed system is $h_{0}(I) + \epsilon^2 f (I, \theta) $. We will now analyze this system without making any specific assumptions about $f(I, \theta)$, which in our system gives rise to source terms seen, for example, in (\ref{dp17}). $f(I, \theta)$ is in general not integrable, but we can apply procedures in Hamiltonian perturbation theory (HPT) to this near-integrable Hamiltonian, to find a canonical transformation that makes it approximately integrable to some order in $\epsilon$. This ensures dynamics still takes place on some possibly perturbed version of the $n$ torus of the linear system ---we characterize this as stability of the system. However, as is known, there is an obstruction to finding such a canonical transformation due to the small denominator problem in HPT. This implies that if the frequencies of the linear oscillators are resonant, with ratios of frequencies rational, even a small perturbation can destroy the corresponding tori of the linear system and dynamics is not restricted any more to be on them. As the system samples more of the phase space, we expect energy to be transferred to a large number of modes, via resonances. In $AdS$, the frequencies of the linear modes are resonant. Therefore, this is another way of viewing the resonant instability of $AdS$, as an instability in \emph{phase} space.

In fact, frequencies which are near-resonant (ratios close to rational) can also lead to the small denominator problem and lack of integrability. If this occurs, a consequence will again be destruction of tori and a nonlinear transfer of energy across modes, leading to an instability. Such frequencies do occur in the gravity-scalar field system and instabilites have been observed numerically in such cases, as further evidence for our picture.
One example is the analysis of a massless scalar field in a cavity in Minkowski spacetime with Neumann boundary conditions on the boundary of the cavity \cite{maliborski2}. An instability was found numerically in this case. The spectrum is not resonant, but for large frequencies, becomes asymptotically resonant (ratios of frequencies become close to rational for higher modes). Further, it has been observed that a certain minimum amplitude is required to trigger instability in this case, as opposed to arbitrarily small perturbations in the fully resonant case \cite{MR}. Massive fields in a cavity in Minkowski spacetime (with either Dirichlet or Neumann boundary conditions) have a nonresonant spectrum that is asymptotically resonant. Instability has been found numerically in this case for small field amplitudes by Okawa, Cardoso and Pani \cite{pani}. These results have been puzzling so far due to the view that instability is associated with a resonant spectrum. Viewing the instability as destruction of tori in phase space provides a reason for instability even when spectra are only asymptotically resonant. We will show further that it also explains the \emph{details} and nuances of numerically observed features.

\subsection{Applying the Benettin-Gallavotti theorem to understand numerical results}
We will now apply a theorem by Benettin and Gallavotti \cite{bg} in nonlinear dynamics (KAM theory), to the system of $n$ perturbed linear harmonic oscillators with hamiltonian $H(I, \theta) = h_{0}(I) + \epsilon^2 f (I, \theta) $. This represents weakly nonlinear perturbations of the gravity-scalar field system. This theorem gives the dynamics of the perturbed system over long times which are an exponential of a power of $\frac{1}{\epsilon^2}$, far longer than the time scales $\sim \frac{1}{\epsilon^2}$ studied in the two-time formalism. We would like to understand the reasons for features observed numerically in the case of asymptotically resonant spectra.

Let $\omega_{i}$, $i=1,...,n$ denote the frequencies of the linear oscillators (modes of the scalar field in the linearized analysis) and let $\pmb{\omega} \in R^{n}$ denote the frequency vector with components $\omega_i $. Let $\textbf{k} \in Z^{n}- \{0\}$ denote a vector of integers. The condition for a resonance is that $\pmb{\omega} \cdot \textbf{k} = 0$ for some $\textbf{k} \in Z^{n}-\{0\}$. If the entire spectrum or part of it is perfectly resonant, we can find $n$ linearly independent $\textbf{k}$ satisfying the resonance condition in the fully resonant case, and $n_1 < n$ in the case where a subspace of the frequency space of dimension $n_1$ is resonant. Let us now consider a spectrum that is not perfectly resonant (either fully or partly). This can be quantified by a \emph{Diophantine condition} on $\pmb{\omega}$ for all $k \in Z^{n} - \{0\}$ and some $\gamma > 0$, namely \footnote{For a discussion of the Diophantine condition, see p.217, \cite{wiggins}.}:
\begin{equation}
| \pmb{\omega} \cdot \textbf{k} | \geq \frac{\gamma}{|k|^{n}};
\label{bg1}
\end{equation}
$|k|$ denoting supremum (over $i$) of $|k_{i}|$. Clearly by choosing large integers for $k_{i}$ we can get arbitrarily close to the resonance condition. $\gamma$ quantifies how close the frequencies are to being perfectly resonant --- asymptotically resonant or nearly resonant spectra will satisfy this condition only for small $\gamma$.

\textbf{Benettin-Gallavotti theorem:}\\
Consider $H(I, \theta) = h_{0}(I) + \epsilon^2 f (I, \theta) $.
Let the frequencies obey the Diophantine condition (\ref{bg1}) for some $\gamma$. Further, denote by $1/E$ the timescale for oscillation of the linearized system, $E = ||\pmb{\omega}||$.  With some further analyticity conditions on $f(I, \theta)$, Benettin and Gallavotti showed the following:

For initial data $(I(0), \theta(0))$, and for amplitude of the perturbation $\epsilon^2 < \epsilon_{0}^{2} $, where $\epsilon_{0}^{2} =  (\frac{D}{E^2 }) \gamma$ and $D>0$ is a constant determined by $n$ and analyticity parameters of $f$, \cite{bg}, \cite{gallavottileshouches};
\begin{equation}
||I(t) - I(0)|| < C (\frac{\epsilon}{\epsilon_{0}});
\label{bg2}
\end{equation}
valid for times $t$ such that
\begin{equation}
t \leq \frac{1}{\epsilon E}(\frac{\epsilon^2}{\epsilon_{0}^{2}})^{-(1/\epsilon^{2} )^{b}}.
\label{bg3}
\end{equation}
where $C$ is a constant. $b = \frac{1}{4(n+1)}$.

In the absence of a perturbation, the action variables are constant, and (\ref{bg2}) indicates that in the presence of the perturbation, the action variables remain close to the initial values for exponentially long times in comparison to the time scale of oscillations of the linear system $\frac{1}{E}$. This is thus an exponential stability result in the case when the frequencies are nonresonant. It is valid for a perturbation of size $\epsilon^2 < \epsilon_{0}^{2}$ where $\epsilon_0$ depends on $\gamma$ in (\ref{bg1}). The stability result will be of practical use if the frequencies are \emph{strongly nonresonant}, i.e., $\gamma$ is not very small. It is important to note that when the condition $\epsilon^2 < \epsilon_{0}^{2}$ is not fulfilled, it does not imply automatically that there is an instability. Rather, the above stability result does not hold and due to the small denominator problem, an instability is very likely. We will discuss the caveats to this statement in II.5.

If the frequencies are asymptotically resonant (i.e., approaching a resonant spectrum as $j \to \infty$), $\gamma$ will be very small, and so will $\epsilon_{0}^{2}$. When the spectrum is asymptotically resonant, the instability, if present, will need a minimum \emph{nonzero} amplitude of the order of $\epsilon_{0}^{2}$ to trigger it, in contrast to what was observed in the fully resonant case. However, that minimum amplitude required for triggering it can be very small if the initial condition has higher frequency mode components for which $\gamma$ is small (leading to small $\epsilon_{0}^{2}$).\footnote{It has also been argued in \cite{horowitz} that higher modes need to be excited for an instability in this case.} Also, the weight of the higher modes with frequency $\omega_j$ in the initial condition matters. We must have the effective amplitude of the higher modes of the field greater than $\epsilon_{0}^{2}$ for triggering instability. If the perturbation has this minimum amplitude, we expect the time scales for the instability in the asymptotically resonant case to be similar to that in the fully resonant case as both are a result of lack of (approximate) integrability of the perturbed system.  This explains the numerical observations in \cite{maliborski2}, \cite{MR} and \cite{pani}.
For a massless scalar field in a cavity of radius $R$ in Minkowski spacetime analyzed by Maliborski \cite{maliborski2}, with Dirichlet boundary conditions on the cavity, the spectrum is resonant. However, with Neumann boundary conditions, an asymptotically resonant spectrum is obtained where $\omega_j $ are the roots of the equation $\tan (\omega_j R) = \omega_j R$. For large $j$, $\omega_j \sim \frac{\pi}{R} (j + \frac{1}{2}) - \frac{1}{2\pi R} (j + \frac{1}{2})^{-1} + O(\frac{1}{j^3})$. Even for $j \sim 100$, retaining only the first (resonant) piece is a good approximation. This implies that if the initial condition has modes of this order, by our discussion, the minimum amplitude required for instability will be small.
Okawa, Cardoso and Pani observe \cite{pani} that even with Dirichlet boundary conditions in the cavity, there is a nonresonant spectrum if the scalar field is \emph{massive}. The frequencies are given by $\omega_j = \sqrt{\mu^2 + (\frac{j \pi}{R})^2}$ where $\mu^2$ is the scalar field mass. This is an asymptotically resonant spectrum when $\frac{\mu^2 R^2}{j^2 \pi^2} <<1$. Then, for sufficiently large $j$ satisfying this inequality, we will have $\omega_j \sim \frac{j \pi}{R} + \frac{\mu^2 R^2}{2j \pi} + O(\frac{1}{j^3}) $. Even if $\mu R = 10$ (at the higher end of the range considered in \cite{pani}), for $j \sim 100$, retaining the first (resonant) term is a good approximation. If the initial condition has modes of this order or higher, an instability can occur with small amplitudes of the field, as shown in \cite{pani}. The normalization used in \cite{pani} was $G=1$, and different from that used in our paper. Keeping $\kappa=8\pi G $ explicitly in the equations only implies that the magnitude of the perturbation term in the Hamiltonian is $\frac{\kappa}{2}~\epsilon^2 $, and $\kappa = 8\pi$ in \cite{pani}.

A review of analytic aspects of the $AdS$ instability, including our work on applications of nonlinear dynamics results to asymptotically resonant spectra can be found in \cite{evninrecent}.

A note of caution is in order here. These results have been arrived at for perturbations of a finite number $n$ of linear oscillators. Simply taking $n \to \infty$ is physically meaningless, because all oscillators are not on an equal footing. In fact, numerics shows a decaying power law for energy per mode with mode frequency. There are infinite dimensional versions of KAM theory, almost all of which have assumptions such that effectively only a finite number of degrees of freedom are significant and the results are similar to what we have mentioned.

\subsection{Fields in (locally) asymptotically $AdS$ spacetimes --- the $AdS$ soliton}

We now illustrate, by means of an example --- the $AdS$ soliton --- the application of the framework of the previous subsections to many (locally) asymptotically $AdS$ spacetimes. Let us consider, for simplicity, the gravity- massless scalar field system in weakly nonlinear perturbation theory where
the scalar field and the metric are expanded as:\\
\begin{equation}
g_{ij} = g_{(0)ij} + \epsilon^2 g_{(2)ij} +....~~~~;~~
\Phi = \epsilon \Phi_1 + \epsilon^3 \Phi_3 +.....
\label{soliton1}
\end{equation}

We now consider the case when $g_{(0)ij}$ is a locally asymptotically $AdS$ metric which is static, diagonal, and with no horizons or singularities. An example of this is the $AdS$ soliton metric (written in five dimensions):
\begin{equation}
ds^2=\frac{r^2}{l^2}\left[\left(1-\frac{r_0^{4}}{r^{4}}\right)d\tau^2+(dx^1)^2 + (dx^2)^2 -dt^2 \right]+\left(1-\frac{r_0^{4}}{r^{4}}\right)^{-1}\frac{l^2}{r^2}dr^2 . \label{soliton2}\end{equation}

The coordinate $r$ is restricted to $r\geq  r_0$ and $\tau$ is a periodic coordinate
which must have a period  $\beta = \frac{\pi l^2}{r_0}$  to avoid a conical singularity at $ r = r_0$. For the detailed construction of this metric, we refer to \cite{hm}.

The Klein-Gordon equation for $\Phi_1$ in a static, diagonal background $g_{(0)ij}$, reduces to the form $$\frac{\partial^2 \Phi_1}{\partial t^2 } + L \Phi_1 = 0;$$ and can be analyzed by separation of variables into modes. For the $AdS$ soliton, let the scalar field depend only on $r$ and $t$. Then  $\Phi_1=\phi_1 (r) \exp(i \omega t)$ and  $L=-\frac{1}{rl^4}\partial_r [(r^5- r_0^4r )\partial_r ]$. The mode frequencies obey the eigenvalue equation $L\phi_1 = \omega^2 \phi_1$. This equation cannot be solved exactly, but the large mode frequencies have been obtained in various ways, either numerically or in a  WKB approximation \cite{csaki}, \cite{jevicki}, \cite{minahan}, \cite{russo}, \cite{myers}. Here, we give the WKB result \cite{myers} valid for $l^4\omega^2/r_{0}^{2} > 5$ (we have set the dependence of their ansatz  on coordinates $x^1$ and $x^2 $ to zero, and their calculation of glueball mass spectra then gives the squared frequency):
\begin{equation}\omega_n^2\simeq n(n+1)\frac{56.67}{\beta^2}+O(n^0) ;\label{soliton2a}\end{equation}
where $n$ are large positive integers. We observe that this is an asymptotically resonant spectrum as for large $n$, $\omega \sim C n$ where $C$ is a constant. The normalizable eigenfunctions corresponding to these frequencies can be evaluated asymptotically, and vanish at the boundary. They are also regular at $r=R$.

Now, for the gravity-scalar field system, with $\Phi(r,t)$, let us consider metrics $g_{ij}$ in the class
\begin{equation}ds^2=-e^{2a(t,r)}\frac{r^2}{l^2}dt^2+e^{2b(t,r)}\frac{l^2 }{f(r)}dr^2+e^{2c(t,r)}\frac{f(r)}{l^2}d\tau^2+e^{2d(t,r)}\frac{r^2}{l^2}\left[(dx^1)^2+(dx^2)^2\right]\label{soliton3}\end{equation}
where $f(r)=r^2(1-\frac{r_0^4}{r^4})$.

This class preserves the planar and circular symmetries of the background. It has been considered in the context of nonlinear gravitational perturbations of the soliton in \cite{kang}.
In Appendix A, we give the full Einstein-scalar field equations for this system.
The $AdS$ soliton background corresponds to the functions $a,b,c,d = 0$. (\ref{soliton1}) gives the expansion for $\Phi$ as well as the metric functions as $$a=a_2 \epsilon^2+a_4\epsilon^4......$$ and similarly for $b,c,d$.
Considering the equations in Appendix A at order $\epsilon^2 $ yields coupled equations for the functions $a_2, b_2, c_2, d_2$. Interestingly, at order $\epsilon^3 $ we get
\begin{eqnarray} &&-rl^4\partial_t\partial_t\Phi_3 +(5r^4-r_0^4)\partial_r\Phi_3
+(r^5-r_0^4r)\partial_r\partial_r\Phi_3 = \nonumber \\&& rl^4(-\dot{a_2}+\dot{b_2}+\dot{c_2}+2\dot{d_2})\partial_t\Phi_1 + rl^4(-2a_2+2b_2)\partial_t\partial_t\Phi_1 \nonumber \\ &&-(r^5-r_0^4r)(a_2'-b_2'+c_2'+2d_2') \partial_r \Phi_1 \label{s23}\end{eqnarray}
This equation can be cast in a form very similar to what was obtained in (\ref{dp16a}) for $AdS$;
\begin{equation}
\partial_t\partial_t\Phi_3+L\Phi_3=S(a_2,b_2,c_2,d_2,\Phi_1)
\label{s24}\end{equation}
where
$L$ is the operator governing the linear modes and $S=-(-\dot{a_2}+\dot{b_2}+\dot{c_2}+2\dot{d_2})\partial_t\Phi_1+\frac{(r^4-r^4_0)}{l^4}(a_2'-b_2'+c_2'+2d_2')\partial_r\Phi_1+2(a_2-b_2)\partial
_t\partial_t\Phi_1$.\\

The simple structure of the equations order by order in $\epsilon$ observed in the scalar field -$AdS$ system is reproduced for this example as well, by considering appropriate classes of perturbations. We can stretch the analogy further by considering the class of reduced perturbations $a=b=0$. For this class, the second order equations can be decoupled to obtain $c_2 $ and $d_2 $ explicitly analogous to (\ref{dp16}) in the $AdS$ case. The details of this calculation are given in Appendix A. If the eigenfunctions of the linear operator $L$ provide a basis for expanding $\Phi_3$, then by projecting (\ref{s24}) onto this basis, we obtain the perturbed linear harmonic oscillator structure discussed extensively in this section. Of course, a rigorous existence/uniqueness analysis for the Einstein-scalar field equations for this class of perturbations as done for $AdS$ in \cite{holzegel} would be complicated in this case. Thus we do not know if the eigenfunctions of $L$ are the appropriate basis, order by order, for $\Phi$. Nevertheless, this exercise shows that it is possible to apply the entire analysis of the previous subsections to static, diagonal, asymptotically $AdS$ spaces, with suitably restricted perturbations. In particular, if the eigenfunctions of $L$ are an appropriate basis, then the asymptotically resonant spectrum of the $AdS$ soliton implies that a necessary condition for instability is presence of the higher modes in the initial condition and a perturbation of a minimum nonzero amplitude. As is well-known, the $AdS$ soliton is the minimum of energy in its asymptotic class \cite{hm}. The instability, if it exists, would be for (finite) scalar field excitations of this background.

We note that a recent paper considering scalar field perturbations of quasi-periodic solutions in $AdS$ \cite{lehnerrecent} has also obtained an asymptotically resonant spectrum for the linear modes of the field.

\subsection{Discussion on further nonlinear aspects}

The KAM (Kolmogorov-Arnold-Moser) theorem \cite{KAM} is often quoted in the $AdS$ instability literature as relevant for long-time behaviour of the gravity-scalar field system. Unlike the Benettin-Gallavotti result (on stability for exponentially long times), with the same assumptions, the KAM theorem guarantees stability as $t \to \infty$ provided the unperturbed Hamiltonian $h_{0}(I)$ obeys the non-degeneracy condition $det[\frac{\partial^2 h_0}{\partial I^2 }] \neq 0$. A set of linear harmonic oscillators do not obey this condition as this determinant is zero. Therefore, we do not see any straightforward application of this theorem, unless the perturbed Hamiltonian can be rewritten as a nonlinear Hamiltonian obeying this, plus a small perturbation. The Nekhoroshev theorem \cite{Nekhoroshev} which shows exponential stability even in the case of a resonant spectrum, with some assumptions (a geometric condition on the Hamiltonian, and the nondegeneracy condition) also does not seem to apply here.

The striking similarity of the $AdS$ gravity scalar field system to coupled nonlinear oscillator Hamiltonian problems such as the FPUT problem \cite{FPU} has been used in \cite{balasubramanian} to uncover quasi-periodic behaviour at times of the order $\frac{1}{\epsilon^2}$. Quasiperiodic solutions in various contexts in the gravity-scalar field system have been  found by \cite{maliborski}, \cite{lehner}, \cite{buchel1}. Selection rules that restrict energy transfer to higher modes (and `inverse energy cascades') have been found in \cite{craps}, \cite{craps1}, \cite{lehner} and more recently, \cite{bizon2}. Similar quasiperiodic behaviour/selection rules have been seen in a system of gravity and self-interacting scalar fields \cite{basu}, \cite{basu1}, \cite{yang}, \cite{krishnan}. At first glance, this is puzzling, since we generically expect a resonant instability in $AdS$, due to its spectrum.

We conjecture that the reason for quasiperiodic behaviour at times of the order $\frac{1}{\epsilon^2}$ could be the `closeness' of the nonlinear scalar field-gravity system to a nontrivial integrable model (i.e., not just a set of linear oscillators) to some order in $\epsilon$. The FPUT system of coupled nonlinear oscillators can also be viewed as a perturbation of linear oscillators. However, this does not explain its complex behaviour. It has been shown that the FPUT model displays quasiperiodic behaviour for certain time scales which can be studied in a two-time framework (such as that employed in some of the papers mentioned). At one scale, FPUT dynamics is indistinguishable from the integrable Toda model \cite{ferguson}, \cite{benettin1} (see also \cite{zabusky} for the connection of the FPUT model to the KdV equation). In the Toda model, coefficients corresponding to nonlinear mode mixing are simply zero, thus those resonant instability channels are absent. At the second time scale, the FPUT model can be regarded as a perturbation of the integrable Toda model (rather than as a perturbation of linear harmonic oscillators). This accounts for the quasi-periodic behaviour for long times (after which one may expect a departure from Toda-like behaviour). The relation of the FPUT model to the KdV equation \cite{zabusky} also suggests that these quasiperiodic solutions are the analogues of the KdV breathers. We conjecture that a similar explanation holds for the quasi-periodic behaviour of the gravity-scalar field system. There may be several ways of writing the perturbed $H(I, \theta) = h_{0}(I) + \epsilon^2 f (I, \theta) $. We have taken $h_{0}(I)$ to be the Hamiltonian for a system of linear harmonic oscillators, but if there is an integrable model with Hamiltonian $\tilde h_{0}(I)$ `close' to our nonlinear system,  $H(I, \theta) = \tilde h_{0} (I) + \epsilon^2 \tilde f (I, \theta)$ could be viewed as a perturbed integrable model. The identification of the integrable model $\tilde h_{0}(I)$ is a hard problem, but one that we expect will open up exciting avenues of future study.

\section{Asymptotics of Jacobi polynomials and significance for localization of field in space}
In the last section, we have seen some necessary conditions for resonant transfer of energy to higher frequencies. However, even if transfer of energy takes place to higher modes, we will still need sufficient localization in space for black hole formation. The key to understanding localization in the $AdS$-scalar field system is the asymptotics of the linear eigenfunctions for high frequency. We argue that given a resonant instability in the gravity-field system in a spacetime, localization in space of high frequency linear modes is a necessary condition for black hole formation, at least when field amplitudes are small.

In studies of the $AdS$ instability, most numerical simulations show a decaying power law behaviour of energy per mode with mode frequency \cite{lecture}. However, at sufficiently large times (close to black hole formation), it is reasonable to expect that there would be significant energy transfer to some high frequency modes. We will discuss the asymptotic form of the modes of the scalar field in $AdS$ for high frequency in various dimensions. In fact, we have checked that the asymptotic (Mehler-Heine) formula we will discuss in this section for the value of the Jacobi polynomial at the maximum $x=0$ only differs by about 1\% from its actual value for mode number $j=100$. In $AdS_4 $, even for $j=10$ there is only an error of about 4\%. The results of this section will be valid for mode numbers in this range as long as sufficient energy transfer has taken place to them. We expect their spatial behaviour to be useful in investigating the localization of the field in space.

The normalized eigenfunctions of the scalar field in a fixed $AdS_{d+1}$ background are given by
\begin{equation}
e_{j}(x) = \frac{2 \sqrt{j!(j+d-1)!}}{\Gamma(j + d/2)} (\cos x )^{d} P_{j}^{(\frac{d}{2} - 1, \frac{d}{2})} (\cos 2x ) ;
\label{jp1}
\end{equation}
where
\begin{equation}
\int_{0}^{\pi/2} e_{j}(x)
e_{l}(x) (\tan x )^{d-1} ~dx = \delta_{jl}.
\label{jp2}
\end{equation}
Now, we want the asymptotic behaviour of $e_{j}(x)$ for large $j$. There are two regimes, one when $x \in (0, \pi/2 )$ and the other at end-points $x=0$ and $x=\pi/2$ .
First, let us consider $x \in (0, \pi/2 )$. For large $j$, from the Darboux formula, we have \cite{askey}
\begin{equation}
P_{j}^{(\frac{d}{2} - 1, \frac{d}{2} )} (\cos 2x ) \sim A(x) \frac{(\frac{1}{2})_j }{j!} 2^{d} \cos (2 j x + \phi ).
\label{jp3}
\end{equation}
Here, $\phi= - (d-1) \frac{\pi}{4}$ and $(p)_{j} = p (p+1)...(p+j-1)$ refers to the Pochhammer symbol.
$$ A(x) = 2^{-d} (\sin x )^{-(d-1)/2}(\cos x )^{-(d+1)/2}.$$
Using this in (\ref{jp1}) and using Stirling's formula, we obtain that for large $j$,
\begin{equation}
e_{j}(x) = \frac{2}{\sqrt{\pi}} (\cos x)^{d} \frac{[\cos (2jx) \cos \phi - \sin (2jx) \sin \phi ]}{(\sin x )^{(d-1)/2}(\cos x )^{(d+1)/2}}.
\label{jp4}
\end{equation}
This formula is not valid at $x=0$ and $x=\pi/2$, however we can certainly consider $x \to \pi/2$, and $e_{j}(x)$ approaches zero in this limit for all $d \geq 2$. Similarly, let us consider the regime when $j$ is large and $x$ small, such that $jx$ is large. In this regime, behaviour varies with dimension:\\
(i) $AdS_3$, or $d=2$: $\phi = - \frac{\pi}{4}$. Clearly, for small $x$,  $\frac{1}{\sqrt{x}}$ is the envelope function of rapidly oscillating $e_{j}(x)$. The envelope function increases slowly as $x \to 0$. For small $x$, $$e_{j}(x) \sim \frac{[C_1 \cos (2jx) - C_2 \sin (2jx) ]}{\sqrt{x}}.$$\\
(ii) $AdS_4$, or $d=3$: $\phi = - \pi/2$. For small $x$, $e_{j}(x) \sim C \frac{\sin (2jx) }{\pi x}$. We know that this function localizes near $x=0$ and is in fact proportional to the delta function $\delta(x)$ as $j \to \infty$.\\
(iii) For $AdS_{d+1}$ with $d>3$, the enveloping function of $e_{j}(x)$ is $\frac{1}{x^{(d-1)/2}}$ as $x \to 0$ and therefore localizes even better near $x=0$.

We now inspect the values of the normalized eigenfunctions at $x=0$ using the Mehler-Heine formula for Jacobi polynomials, which gives their values in a neighbourhood of $x=0$ and $x=\pi/2$ for large $j$ \cite{askey}. From the formula, we can easily check that $e_{j}(\pi/2) = 0$.
Define $z$ by $\frac{z}{j} =2x$. Then, in a neighbourhood of $x=0$,
\begin{equation}
\lim_{j \to \infty} j^{-\alpha} P_{j}^{\alpha,\beta} (\cos \frac{z}{j} ) = (z/2)^{-\alpha} J_{\alpha}(z).
\label{jp5}
\end{equation}
$J_{\alpha}(z)$ is the Bessel function. In the regime where $j$ is large, $x$ is small, such that $z$ is large, the Mehler-Heine formula gives the same asymptotics as the Darboux formula (\ref{jp4}).

If $j$ is large and $x \to 0$, such that $z \to 0$, then $J_{\alpha}(z) \sim \frac{1}{\Gamma(\alpha + 1)} (z/2)^{\alpha}$. Thus, for $x = 0$, $P_{j}^{\alpha,\beta} (1) \sim \frac{1}{\Gamma(\alpha + 1)} j^{\alpha}$. Here,
$\alpha = \frac{d}{2} - 1 $. This can also be seen from the relation of the Jacobi polynomial to the hypergeometric function.
We next want $e_{j}(0)$. From (\ref{jp1}), we need to estimate the large $j$ behaviour of the normalization factor $\frac{2 \sqrt{j!(j+d-1)!}}{\Gamma(j + d/2)}$. Using Stirling's formula for the Gamma function, this is $\sqrt{\frac{j}{e}}$.
Thus $e_{j}(0) \sim \frac{1}{\Gamma(\alpha + 1)} j^{\frac{d-1}{2}}$. This increases as $j$ increases for all $d \geq 2$, with the increase being the smallest for $d=2$, i.e., $AdS_3$.

We thus conclude that localization of the eigenfunctions around $x=0$ for large $j$ increases with dimension $d$.

Thus if energy is transferred to higher frequency modes for which the asymptotic Darboux and Mehler-Heine formulae apply, their spatial behaviour and localization properties lead to localization of the field, at least for a wide variety of initial conditions  with a single pulse such as the Gaussian in \cite{bizon}. We see that localization increases with dimension, and is least for $d=2$ ($AdS_3$). In $AdS_3$, numerical studies show a very fast transfer of energy to higher frequencies (in comparison to higher dimensions), yet no black hole formation \cite{jalmuzna1}, \cite{jalmuzna2}. We speculate that at least part of the reason may be that the eigenfunctions localize the least in three dimensions. For asymptotically $AdS$ spacetimes, a similar study of the eigenmodes at high frequency needs to be done to see if localization happens. Localization of eigenfunctions for high frequencies is a necessary (but not sufficient) condition for an instability to result in black hole formation for a wide variety of initial conditions.

Curiously, the asymptotic Darboux formula for the scalar field mode in $AdS_4 $ is similar to the exact form of the mode eigenfunction for the scalar field in a cavity of radius $R$ in $4$-dimensional Minkowski spacetime with Dirichlet boundary conditions. Both when the scalar field is massless \cite{maliborski2} and when it is massive \cite{pani}, the eigenfunctions corresponding to the linear modes are $ e_{j} (r) = \sqrt{\frac{2}{R}} \frac{\sin (\omega_j r)}{r}$. Clearly for $\omega_j $ large, these eigenfunctions localize (approaching a delta function in the limit $\omega_j \to \infty$). There is an instability in this system leading to horizon formation, consistent with our arguments.

We would like to end the section with a word of caution. If one were to start with an initial condition with multiple pulses at different locations, then the interactions between them could be very complicated \cite{freivogel}. In such a scenario, one cannot conclude naively about localization from the eigenfunctions alone. One could also have initial data comprising of a few modes, but with an amplitude so large that black hole formation occurs without the need for resonant transfer of energy. Further, in gravity with a Gauss-Bonnet term \cite{kunstatter}, this term results (in some cases) in a pulse breaking up into multiple pulses, with relative phases between themselves, and again asymptotics of eigenfunctions may not be useful.
Asymptotics of eigenfunctions for $x \neq 0$ have also been used recently in a different context, to compute interaction coefficients between modes in \cite{evnin1}, \cite{evnin}.

\section{Summary}
The $AdS$ instability makes it imperative to identify the necessary conditions for instabilities to be caused by nonlinearities in the gravity-field system. In particular, one would like to know when arbitrarily small field perturbations can cause an instability. This is difficult to probe numerically and an analytical understanding is needed.

In this paper, we provide, for the first time, an analytical framework to identify the necessary conditions for an $AdS$-type instability in a gravity-scalar field system. It also predicts the late-time behaviour of arbitrarily small $AdS$ perturbations. In this framework, the $AdS$ instability is understood as a dynamical instability of the scalar field in phase space. Dynamics in the linearized approximation is integrable, and nonlinearities can lead to breakdown of integrability. If this happens, the scalar field can sample a higher-dimensional subset of phase space than the integrable case, with the only restriction coming from energy conservation. This manifests as nonlinear energy transfer across modes, leading to an instability.

We apply a theorem in nonlinear dynamics, from KAM theory, to derive the necessary condition for such an instability.  The \emph{spectrum} of the linear modes of the field is crucial. If the spectrum is strongly nonresonant (i.e, ratios of frequencies are irrational numbers badly approximated by rational numbers in a precise sense), the gravity-scalar field system is stable for long times (even with reflecting boundary conditions). Thus, a necessary condition for instability is a spectrum that is resonant or approximately/asymptotically resonant. This is not sufficient, but an instability is likely in this case. The reason is a breakdown of integrability in the nonlinear system due to the small denominator problem in nonlinear dynamics. A perfectly resonant spectrum, in general, leads to a resonant instability for arbitrarily small field amplitudes. If the spectrum is asymptotically resonant, we show that such modes can trigger an instability if they are excited by the initial condition, and if the perturbation has a certain minimum nonzero amplitude (whose value depends in a precise sense on the deviation of the spectrum from perfect resonance). This explains puzzling numerical results for a scalar field with an asymptotically resonant spectrum in a cavity in Minkowski spacetime \cite{maliborski2}, \cite{MR}, \cite{pani}.

Our framework is applicable to (scalar, gravitational and other) fields in many asymptotically $AdS$ spacetimes for a wide class of non-dissipative (self-adjoint) boundary conditions. As an example, we consider the $AdS$ soliton-scalar field system in five dimensions, and identify a class of perturbations for which the structure of the equations bears a close resemblance to the $AdS$ case. Since the spectrum of the linear modes of the field in the $AdS$ soliton background is known to be asymptotically resonant, we predict the necessary condition for an instability leading to an energy cascade to higher modes.

The outcome of a resonant instability in a gravitational system need not be a black hole. The necessary (but not sufficient) condition for that to happen given a resonant instability is a strong localization of modes of the linear system for higher frequencies. We have argued that in $AdS_{d+1}$, this happens for a wide range of initial conditions due to the asymptotics and localization properties of Jacobi polynomials. Localization is least in $AdS_3$. This provides the best explanation for lack of black hole formation in $AdS_3$ despite efficient energy transfer to higher modes.

These results together allow us to address the issue of stability of asymptotically $AdS$ spacetimes with respect to scalar fields and gravitational perturbations (which have been discussed in \cite{horowitz1}, \cite{horowitz}). The next direction is a systematic study of asymptotically $AdS$ spacetimes both with respect to their linear spectrum (under scalar and gravitational perturbations) and the localization properties of their eigenfunctions. This relatively simple exercise will tell us about their stability or the outcome of an instability. 
In the next section and in Appendix B, we have also offered a reason for surprising differences in the gravity-scalar field system between Einstein gravity and Einstein-Gauss-Bonnet gravity observed in \cite{kunstatter}.

\section{Discussion, future directions}

As discussed in section II, fascinating quasi-periodic behaviour has been observed in numerical studies of the $AdS$ instability problem until times $t \sim \frac{1}{\epsilon^2 }$ for some initial conditions. We have conjectured, based on our framework, that the reason could be proximity of the gravity-scalar field Hamiltonian to a nonlinear integrable Hamiltonian to $O(\epsilon^2 )$. This is similar to the Fermi-Pasta-Ulam system, whose quasi-periodic behaviour is explained by the fact that it is a perturbation of the integrable Toda model. A very interesting problem is the identification of the nonlinear integrable Hamiltonian that is close to the gravity-scalar field Hamiltonian to $O(\epsilon^2 )$. The various selection rules and results from the two-time formalism for studying the $AdS$ instability will be crucial to solving this problem.

One class of spacetimes to which our results do not apply are spacetimes with horizons, such as $AdS$ black holes. This is due to the fact that quasinormal mode boundary conditions on perturbations at the horizon are dissipative, and the results we have used from nonlinear dynamics do not apply to systems with dissipation. The issue of nonlinear stability for some rotating black holes under gravitational perturbations has been discussed in \cite{horowitz}. However, it is known that even in the linearized approximation, small Kerr-$AdS$ black holes are unstable due to superradiance \cite{kerradscardoso1}, \cite{kerradscardoso2}. Numerical work on linear evolution of scalar fields in small $AdS$ black hole spacetimes in the time domain confirms this \cite{cardosotimedomain}. It is very important to understand how the nonlinear effects impact the end-point of this instability. Recently, the set-up for full numerical nonlinear evolution of a scalar field in asymptotically $AdS$ black hole spacetimes has been developed \cite{pretorius}. The results from this program may lead to insights on the role of nonlinearities in black hole perturbations. We are actively working, from an analytical point of view, on applications of nonlinear dynamics of dissipative systems to these exciting problems.

A potentially rich and complex topic is the study of nonlinear instabilities in the gravity-scalar field system in theories of gravity with higher curvature corrections. The gravity-scalar field system was investigated recently in five dimensional Gauss-Bonnet gravity \cite{kunstatter} for the same set-up as that of Bizon and Rostworowski (reviewed in II.1). There are surprising differences to Einstein gravity --- in particular, the Gauss-Bonnet term leads to greater dispersion of the scalar field. One observation in \cite{kunstatter} is particularly interesting: for $\epsilon$ (field amplitude) less than a certain nonzero critical value (corresponding to the critical mass for no black hole to form), there is not even a resonant transfer of energy to higher scales. This needs to be analytically understood; however, we offer a reason for why the dynamics with the Gauss-Bonnet term is different from the pure Einstein case. Let us denote the Gauss-Bonnet coupling by $\lambda_3$. The equations of the Einstein-Gauss-Bonnet (EGB) scalar field system  are given by (\ref{gb1}) and (\ref{gb2}) in Appendix B.
Studying the system order by order in weakly nonlinear perturbation theory, there are corrections to the corresponding equations in Einstein gravity beyond the linear order in $\epsilon$. We have found that, by defining a rescaled field $\psi$ by  $$\frac{\phi}{(1-2\lambda_3)^{1/2} }=\psi, $$ the equations for the EGB scalar field system for $\phi$ are identical to the Einstein gravity-scalar field equations up to $O(\epsilon^3)$ for $\psi$. The Gauss-Bonnet term cannot be scaled away at order $\epsilon^4 $ and higher. The detailed equations up to $O(\epsilon^4)$ before and after rescaling can be found in Appendix B.
Interestingly, precisely the same rescaling has been observed before in a two-time 
formalism (which is valid to $O(\epsilon^3)$) 
in \cite{buchellehner}\footnote{Our conventions for the cosmological constant differ from those in this paper, but after 
taking this into account, the rescaling factors are the same.}. The authors of \cite{buchellehner} interpret this
observation holographically using the AdS/CFT correspondence and  
conjecture that nonequilibrium dynamics of the conformal field theory (CFT) dual in the presence of the Gauss-Bonnet term is similar to that
of the CFT in the absence of it up to times $O(1/\epsilon^2 )$. 

Thus, at least at order $\epsilon^3 $, a perturbation in Gauss-Bonnet gravity is dynamically equivalent to a larger 
field perturbation in Einstein gravity --- larger by the factor $\frac{1}{(1-2\lambda_3)^{1/2} }$. This factor can 
be significant if $\lambda_3$ is large. Even if $\epsilon$ were small, this factor being large would 
paradoxically imply that the system is dynamically equivalent to Einstein gravity to order $\epsilon^3 $ but in a 
regime when weakly nonlinear perturbation theory is not expected to be valid. We therefore expect complex behaviour. 
As well, if $\epsilon$ is not very small, clearly the perturbative picture breaks down and the Gauss-Bonnet 
term can play an important role as it introduces new nonlinearities.
The regime that would be interesting to study is one where $\epsilon$ and $\lambda_3 $ are both small. In this regime, 
at every order $\epsilon^4 $ and higher, the terms involving the Gauss-Bonnet coupling $\lambda_3$ are \emph{not} the 
leading terms growing in time. So a priori, it does not seem that there would be drastic differences to 
Einstein gravity until late times when the perturbation has grown, and the Gauss-Bonnet term could dominate. The 
conjecture in \cite{buchellehner} is precisely the holographic dual of this observation.
Indeed, as the authors of \cite{buchellehner} point out, holography constrains the value of $\lambda_3$ to be small, and
this seems to be the physically interesting regime. 
This issue merits further numerical investigation, and a deeper analytical understanding. We hope to report on these exciting problems in the near future.

\section{Acknowledgements} One of us (V S) would like to thank A D Gangal for several enjoyable discussions on KAM theory.
We would like to thank the referee for bringing \cite{buchellehner} to our attention.

\section{Appendix A}
For the class of five dimensional metrics given by (\ref{soliton3}) and a massless scalar field $\Phi (r,t)$,
the Einstein-scalar field equations can be written as ($\kappa = 8\pi G$):
\begin{equation}R_{ab}+\frac{4}{l^2}g_{ab}= W_{ab}\label{s4}\end{equation}

\begin{equation}\frac{1}{\sqrt{-g}}\partial_a(g^{ab}\sqrt{-g}\partial_b\Phi)=0\label{s5}\end{equation}
where $W_{ab}=\kappa(T_{ab}-\frac{1}{3}g_{ab}T)$ and  $
T_{ab}=\partial_a\Phi\partial_b
\Phi-\frac{1}{2}g_{ab}(\partial\Phi)^2
$.
\\
Plugging (\ref{soliton3}) into (\ref{s4}) \& (\ref{s5}) we get
\begin{align}
&\frac{e^{-2b}}{l^4}\Big[ e^{2a} rf'(1+ra')+e^{2a}f\{2+r^2a'^2-rb'+rc'+2rd'+4ra'-r^2a'b' \nonumber \\ &+r^2a'c' +2r^2a'd' +r^2a''\}
+e^{2b}l^4\{\dot{a}\dot{b}+\dot{a}\dot{c}-\dot{b}^2 -\dot{c}^2-2\dot{d}^2 +2\dot{a}\dot{d}-\ddot{b} \nonumber \\ &-\ddot{c}-2\ddot{d}\}\Big]-\frac{4r^2e^{2a}}{l^4}=W_{tt} \label{ee1}
\end{align}

\begin{align}
&\frac{e^{-2a}}{2r^2f}\Big[-e^{2a}r^2f''-e^{2a}rf'(3+ra'-rb'+3rc'+2rd')-2\{e^{2a}rf(ra'^2+2a'\nonumber \\&-ra'b'+rc'^2+4d'+2rd'^2
-3b'-rb'c'-2rb'd'+ra''+rc''+2rd'')+\nonumber \\ &e^{2b}l^4(\dot{a}\dot{b}-\dot{b}\dot{c}-2\dot{b}\dot{d}-\dot{b}^2-\ddot{b})\}\Big]+\frac{4e^{2b}}{f}=W_{rr}   \label{ee2}
\end{align}

\begin{align}
&\frac{-e^{-2a-2b+2c}f}{2l^4r^2} \Big[e^{2a}r^2f''+e^{2a}rf'(3+ra'-rb'+3rc'+2rd')+2\{e^{2a}rf (rc'^2\nonumber \\&
+3c'+ra'c'-rb'c'+2rc'd'+rc'')+e^{2b}l^4(\dot{a}\dot{c}-\dot{b}\dot{c}-2\dot{c}\dot{d}-\dot{c}^2-\ddot{c})  \}\Big]\nonumber \\&+\frac{4e^{2c}f}{l^4}=W_{\tau\tau}\label{ee3}
\end{align}

\begin{align}
&-e^{-2a-2b+2d}\Big[ e^{2a}rf'(1+rd')+e^{2a}f\{2+rc'+5rd'+r^2c'd'+2r^2d'^2+ra'\nonumber \\&+r^2a'd'-rb'-r^2b'd'+r^2d''\} +e^{2b}l^4(\dot{a}\dot{d}-\dot{b}\dot{d}-\dot{c}\dot{d}-2\dot{d}^2-\ddot{d})\Big]\nonumber \\&+4r^2e^{2d}=W_{xx}l^4 \label{ee4}
\end{align}
with $
W_{tt}=\kappa \dot{\Phi}^2,\hspace{1 mm}
W_{rr}=\kappa \Phi'^2\hspace{1 mm}
\& \hspace{1 mm}W_{\tau\tau}=W_{xx}=0$.

\begin{align}&-rl^4 e^{-a+b}\Big(\partial_t\partial_t\Phi+(-\dot{a}+\dot{b}+\dot{c}+2\dot{d})\partial_t\Phi\Big)+e^{a-b}\Big((r^5-r_0r)\partial_r\partial_r\Phi+\nonumber \\&(5r^4-r_0^4)\partial_r\Phi+(r^5-r_0^4r)(a'-b'+c'+2d')\partial_r\Phi\Big)=0
\label{ee5}
\end{align}
Let us now consider a reduced class of perturbations with $a=b=0$.
Putting $a=b=0$ in the equations (\ref{ee1}-\ref{ee4}), one obtains:

\begin{align} \frac{rf'}{l^4}+\frac{2f}{l^4}+\frac{rfc'}{l^4}+\frac{2rfd'}{l^4}-\dot{c}^2-2\dot{d}^2-\ddot{c}-2\ddot{d}-4\frac{r^2}{l^4}= W_{tt} \label{s6} \end{align}

\begin{align}&-\frac{1}{2rf}\Big[rf''+3f'+3rf'c'+2f'rd'+2f(rc'^2+4d'+2rd'^2+rc''+2rd'')\Big]\nonumber \\&+\frac{4}{f}=W_{rr} \label{s7}\end{align}

\begin{align}&\frac{-1}{2r^2}\Big[r^2f''+rf'(3+3rc'+2rd')+2rf(rc'^2+3c'+2rc'd'+rc'')-l^4(2\dot{c}^2\nonumber \\&+4\dot{c}\dot{d}
+2\ddot{c})\Big]+4=0.\label{s8}\end{align}

\begin{align}&-rf'-r^2f'd'-2f-5frd'-2fr^2d'^2-frc'-fr^2c'd'-fr^2d''+l^4(\dot{c}\dot{d} \nonumber \\&+2\dot{d}^2+\ddot{d})+4r^2= 0.\label{s9}\end{align}
Upon eliminating $\ddot{c}$ and $\ddot{d}$ from  these equations, we get\\
\begin{align}
&-\frac{r^2f''}{2}-\frac{5rf'}{2}-4fr(c'+2d')-\frac{3}{2}r^2f'(c'+2d')-4fr^2c'd'+8r^2+2l^4(\dot{d})^2\nonumber \\&-r^2f(c'^2+4d'^2)- fr^2(c''+2d'')
+4l^4\dot{c}\dot{d} - 2 f =l^4W_{tt}\label{s10}\end{align}

\begin{eqnarray}
&&rf''+3f'+3rf'c'+2rf'd'+2fr(c')^2+8fd'+4rf(d')^2+2frc''+4frd''\nonumber \\&&-8r=-2rfW_{rr} \label{s11}\end{eqnarray}
At the second order (\ref{s10}) \& (\ref{s11}) take the form\\
\begin{equation}-4fr(c'_2+2d'_2)-\frac{3}{2}r^2f'(c'_2+2d'_2)-fr^2(c''_2+2d''_2)=\kappa l^4(\dot{\Phi}_1)^2 \label{s19}\end{equation}
\begin{equation}r^2f'(3c_2'+2d_2')+8rfd_2'+2fr^2(c_2''+2d_2'')=-2\kappa r^2f(\Phi'_1)^2 \label{s18}\end{equation}
Let $c_2+2d_2=X$.
Then  (\ref{s19}) becomes\\
\[-4frX'-\frac{3}{2}r^2f'X'-fr^2X''=\kappa l^4(\dot{\Phi}_1)^2\]
\[X''+\frac{4fr+\frac{3}{2}r^2 f'}{fr^2}X'=-\kappa l^4\frac{(\dot{\Phi}_1)^2}{fr^2}\]
\[ X''+ P(r)X'=-\kappa l^4\frac{(\dot{\Phi}_1)^2}{fr^2}  \]
where $P(r)=\frac{4fr+\frac{3}{2}r^2 f'}{fr^2}$.
Therefore,
\begin{equation}X'=-\kappa l^4 e^{-\int^r P(r)dr}\int^r e^{\int^r P(r)dr}\frac{(\dot{\Phi}_1)^2}{fr^2} dr\end{equation}
Thus \\
\begin{equation}c_2'= \frac{1}{4rf-2f'r^2}\left[ 2\kappa (r^2f(\Phi_1')^2-l^4(\dot{\Phi}_1)^2)-X'(4rf+2r^2f')\right] \label{s20}\end{equation}

\begin{equation}d_2'=\frac{X'-c_2'}{2} \label{s21}\end{equation}
By integrating both sides of (\ref{s20}) and (\ref{s21}) we can obtain $c_2$ and $ d_2$ .
Putting $a_2=b_2=0$ in (\ref{s23}) we get the equation governing the dynamics of  the system at the third order, which is :
\begin{equation}
\partial_t\partial_t\Phi_3+L\Phi_3=S(c_2,d_2,\Phi_1)
\label{s24a}\end{equation}
where $S=-(\dot{c_2}+2\dot{d_2})\dot{\Phi}_1+\frac{r^2f}{l^4}(c_2'+2d_2')\Phi_1'$.\\

\section{Appendix B}
The equations of motion for the Einstein-Gauss-Bonnet (EGB) scalar field system are:
\[
G_{ab}+\Lambda g_{ab}+\lambda_3 H_{ab}=8\pi G(\partial_a\phi\partial_b\phi-\frac{1}{2}g_{ab}(\partial\phi)^2)
\]
\begin{equation}
g^{ab}\nabla_a \nabla_b\phi=0
\label{gb1}
\end{equation}
where
\begin{eqnarray}
H_{ab}&=& 2(RR_{ab}-2R_{a\mu}R^\mu_{~b}-2R^{\mu\nu}R_{\mu a\nu b}+R_a^{~\mu\nu\rho}R_{b\mu\nu\rho}) \nonumber \\ & &- \frac{1}{2} g_{ab}(R^2-4R^{\mu\nu}R_{\mu\nu}+R^{\mu\nu\sigma\rho}R_{\mu\nu\sigma\rho})
\label{gb2}
\end{eqnarray}
$\lambda_3$ is the Gauss-Bonnet coupling.
As shown in \cite{kunstatter}, this can be cast in first-order form as in the Einstein gravity-scalar field case. Studying the system order by order in weakly nonlinear perturbation theory as in II.1, there are corrections to the corresponding equations in Einstein gravity beyond the linear order in $\epsilon$. The equations to $O(\epsilon^4 )$ are:\\
At the first order: (the equation is identical to the Einstein gravity case)
\begin{equation}-\ddot{\phi_1}+\frac{1}{\tan^3x}\partial_x(\tan^3x (\partial_x\phi_1))=0\label{gb10}\end{equation}
At second order:\\
\begin{equation}\delta_2=-\frac{1}{(1-2\lambda_3)}\int_0^x\cos y\sin y\big((\phi_1')^2+(\dot{\phi_1})^2\big)dy \label{gb11}\end{equation}

\begin{equation}A_2=\frac{1}{(1-2\lambda_3)}\frac{\cos^4(x)}{\sin^2(x)}\int_0^x \tan^3y\big((\phi_1')^2+(\dot{\phi_1})^2\big)dy \label{gb12}\end{equation}
At the third order:
\begin{equation}-\ddot{\phi_3}+\frac{1}{\tan^3x}\partial_x(\tan^3x (\partial_x\phi_3))=2\ddot{\phi_1}(\delta_2+A_2)+\dot{\phi_1}(\dot{\delta_2}+\dot{A_2})+\phi_1'(A_2'+\delta_2')\label{gb13}\end{equation}

At the fourth order:
\begin{align}A_4&=&
\frac{\cos^4x}{\sin^2x}\Big\{\int_0^x \Big[\frac{-\lambda_3}{(1-2\lambda_3)}\Big(\frac{(A_2)^2}{\cos^4(y)}\Big)' +\frac{1}{(1-2\lambda_3)}\tan^3y(2\phi_1'\phi_3'+2\dot{\phi_1}\dot{\phi_3}  \nonumber\\&&  +2\delta_2(\dot{\phi_1})^2+A_2(\dot{\phi_1})^2-A_2(\phi_1')^2)\Big]dy \Big\}\label{gb14}
\end{align}

\begin{align}\delta_4&=&\int_0^x \Big[-\frac{\cos y\sin y}{(1-2\lambda_3)}\big(2\phi_1'\phi_3'+2\dot{\phi_1}\dot{\phi_3}+2A_2(\dot{\phi_1})^2
+2\delta_2(\dot{\phi_1})^2)\nonumber \\&&-\frac{2\lambda_3}{(1-2\lambda_3)}\frac{\delta_2'A_2}{\sin^2(y)}\Big]dy \label{gb15}\end{align}

Let us now consider the following rescaling of the field:
\begin{equation}\frac{\phi}{(1-2\lambda_3)^{1/2} }=\psi \label{gb16}\end{equation}

At the first order, there is no difference in the equation for $\phi$ and $\psi$.
\begin{equation}-\ddot{\psi_1}+\frac{1}{\tan^3x}\partial_x(\tan^3x (\partial_x\psi_1))=0\label{gb17}\end{equation}
At the second order, this equation is identical to the Einstein gravity-scalar field system for $\psi$.\\
\begin{equation}\delta_2=-\int_0^x\cos y\sin y((\psi_1')^2+(\dot{\psi_1})^2)dy \label{gb18}\end{equation}

\begin{equation}A_2=\frac{\cos^4x}{\sin^2x}\int_0^x \tan^3(y)((\psi_1')^2+(\dot{\psi_1})^2)dy \label{gb19}\end{equation}
At the third order;
\begin{equation}-\ddot{\psi_3}+\frac{1}{\tan^3x}\partial_x(\tan^3x (\partial_x\psi_3))=2\ddot{\psi_1}(\delta_2+A_2)+\dot{\psi_1}(\dot{\delta_2}+\dot{A_2})+\psi_1'(A_2'+\delta_2')\label{gb20}\end{equation}
At the fourth order, there are terms proportional to $\lambda_3 $ that cannot be scaled away.
\begin{align}A_4&=&
\frac{\cos^4x}{\sin^2x}\Big\{\int_0^x \Big[\frac{-\lambda_3}{(1-2\lambda_3)}\Big(\frac{(A_2)^2}{\cos^4(y)}\Big)' +\tan^3y(2\psi_1'\psi_3'+2\dot{\psi_1}\dot{\psi_3}+2\delta_2(\dot{\psi_1})^2\nonumber \\&&
+A_2(\dot{\psi_1})^2-A_2(\psi_1')^2)\Big]dy \Big\}\label{gb21}
\end{align}

\begin{align}\delta_4&=&\int_0^x \Big\{-\cos y\sin y\big(2\psi_1'\psi_3'+2\dot{\psi_1}\dot{\psi_3}+2A_2(\dot{\psi_1})^2
+2\delta_2(\dot{\psi_1})^2)\nonumber\\&&-\frac{2\lambda_3}{(1-2\lambda_3)}\frac{\delta_2'A_2}{\sin^2(y)}\Bigr\}dy \label{gb22}\end{align}

\end{document}